\documentclass[prb,preprint,aps]{revtex4-1}
\usepackage{graphicx}
\usepackage{amsmath}
\usepackage{amssymb}
\usepackage{epsfig}
\usepackage{color}

\begin{document}

\title{Statistical properties of $1D$ spin glasses from first principles of classical mechanics}

\author{ A. S. Gevorkyan$^{1,2}$ and V. V. Sahakyan$^{1}$}

\address{$^{1}$ Institute for Informatics and Automation Problems, NAS of RA}
 \address{$^{2}$ Institute of Chemical Physics, NAS of RA, \\
 g\_\,ashot@sci.am}

\begin{abstract}
We study the classical $1D$ Heisenberg spin glasses, assuming that the orientation of spins
are a spatial. The system of \emph{recurrence equations }(RE) is obtained, by minimization of the
nearest-neighboring Hamiltonian in nodes of $1D$ lattice. It is shown that in each node
of the lattice there is a probability of bifurcation of the solution of REs. This leads to
the fact that, performing a consecutive node-by-node calculations on the $n$-th step instead
of a single stable spin-chain we get a set of spin-chains (strings) which form Fibonacci subtree (graph).
Assessing the computational complexity of one graph  shows that it is $\propto 2^nK_s$, where
$n$ and $K_s$ denote the subtree height  and Kolmogorov's complexity of a string  respectively.
It is shown that the statistical ensemble may be represented as a set of random graphs, where
the computational complexity of each graph is $\mathbb{NP}$ hard.  It is proved, that all
strings of the ensemble have the same weights. This allows in the limit of statistical
equilibrium, with a predetermined accuracy to reduce the initial $\mathbb{NP}$ hard problem
to the $\mathbb{P}$ problem. As shown, the statistical distributions of different
parameters, which are performed by using $\mathbb{NP}$ and $\mathbb{P}$ algorithms for
the respective curves provide a perfect coincidence. Lastly using the formal similarity
between  the ergodic dynamical system and ensemble of spin-chains, a new representation
for the partition function in the form of one dimensional integral from the spin-chains'
energy distribution is proposed.
 \end{abstract}

\maketitle
\section{Introduction}
\label{main}
A wide class of phenomena  in physics, chemistry, material science, biology,
nanoscience, neural network, evolution, organization dynamics, hard-optimization,
environmental and social structures, human logic systems, financial mathematics
etc, mathematically  are well described by models of spin glasses
\cite{Bind,Mezard,Young,Ancona,Bov,Fisch,Tu,Chary,Baake,ash}. Despite numerous
studies nonetheless  there are still a number of topical issues in the field of spin
glasses and  disordered systems as a whole, the solution of which is extremely important
from the point of view of the development of modern technologies. We can mention
important ones of them;

a) The simulation of spin glasses  far from  thermodynamic equilibrium.
It is obvious, in such cases, we can not enter the ambient temperature and, respectively,
write and use a standard representation for partition function.

b) Even if it is assumed that spin glass is in the state of the thermodynamic equilibrium,
and for it may be written  in the standard form the partition function,  in the frameworks of standard theoretical
and numerical methods,it remains an open research question of metastable states. Recall that
the Monte Carlo simulation methods  allow us to study the spin systems only in the \emph{ground state},
at the time when the real statistical system, all the more spin glasses, always are in the metastable
states, i.e in the state where  characterizing the spin glass parameters have some distributions.

c) At definition of the partition function, a priori is assumed that the total weight of
nonphysical spin configurations in the configuration space is a zero that in a number of cases
may be an incorrect assumption. Recall that under the nonphysical spin configurations, we mean
such spin-chains, which are unstable based on the basic principles of classical mechanics.

d) The computational complexity of spin glasses often  applies to the class of the $\mathbb{NP}$ hard
problems. This circumstance to require the development of new efficient algorithms for a numerical
simulation of spin glasses that one way or another leads to the problem of reduction of the
$\mathbb{NP}$ hard to the $\mathbb{P}$ problem.

As it was shown in works \cite{Lie,Ang,CP,HC}, the problem of spin glasses even in the state
of the thermodynamic equilibrium often are $\mathbb{NP}$ hard problems, whose source of which
is in the diverging equilibration at simulations by the Monte Carlo methods \cite{Metropolis}.
In the last time in the statistical physics occurs a rapid growth  the number of works on
methods of the combinatorial optimization \cite{Hayes,Mon,Dub}. In particular
a number of disordered statistical systems have been mapped onto combinatorial
problems, for which a fast combinatorial optimization algorithms are available \cite{Alava,Hart}.
So, combinatorial methods and corresponding algorithms are often used for a simulation of
spin glasses especially when studying the phenomena such as phase transitions where they
have given valuable insights about questions that are hard to investigate by traditional
techniques, for example by Monte Carlo simulations \cite{Lie}. However, the above-mentioned
questions, on which we want to obtain clear answers, obviously, require to development of
principally new approaches.

In this paper we will study the classical 1$D$ spin glass problem suggesting that only the
nearest neighboring spins  interact. Recall despite the simplicity of the model, since in
a known sense it's an exactly solvable model \cite{Tomp}, as it  will shown below, all the
aforementioned problems in this model are present, if we try to solve the task from first
principles of classical mechanics.

One of the important goals of this work is to prove, that in the limit of statistical equilibrium
the initial $\mathbb{NP}$-hard problem with the prescribed accuracy can be reduced to the $\mathbb{P}$
problem, that in turn implies the creation of high-performance algorithm for simulation of
the Heisenberg type spin glasses. In the work  possibilities of generalization of the model for
descriptions of more complex and realistic disordered systems of nature are also discussed.

\section{Definition of model}

The disordered 1$D$ spin-chain in the framework of the nearest-neighboring Heisenberg model
may be written as:
\begin{eqnarray}
H=-\sum_{i\,\in\, \mathcal{N}} J_{i,\,i+1}\textbf{\emph{s}}_i\textbf{\emph{s}}_{i+1},
\qquad \textbf{\emph{s}}_i\in \mathbb{R}^3,\qquad
||\textbf{\emph{s}}_i||=||\textbf{\emph{s}}_{i+1}||=1,
 \label{01}
\end{eqnarray}
where $\mathcal{N}=\{1,...,n\} $  is the set of nodes on $1D$ lattice, the couplings $J_{i,\,i+1}$
are independent random variables characterizing the power of interactions between spins. The
distribution of the coupling constants will be found below as a result of the numerical simulation.

Since the norm of vector $\textbf{\emph{s}}_i=(x_i,y_i,z_i) $ is equal to the unit, then the
projection, $z_i$ can be represented in the following form:
\begin{eqnarray}
\emph{\textbf{z}}_i=q_i|z_i|,\qquad
z_i=(1-x_i^2-y^2_i)^{1/2}>0,\quad q_i=\mathrm{sign}(\emph{\textbf{z}}_i),
 \label{02}
\end{eqnarray}
where $q_i$ is a discrete variable which can take two possible values +1 and -1.

Substituting the Hamiltonian (\ref{01}) into Hamilton equations (see for example \cite{Gold})
can be found:
\begin{eqnarray}
-\ddot{x}_i=J_{i-1,\,i}\bigl(x_{i-1}-{x_i}{z_i}^{-1}z_{i-1}\bigr)+
J_{i,\,i+1}\bigl(x_{i+1}-{x_{i}}{z_{i}}^{-1}z_{i+1}\bigr),
\nonumber\\
-\ddot{y}_i=J_{i-1,\,i}\bigl(y_{i-1}\,-{y_i}{z_{i}}^{-1}z_{i-1}\bigr)+
J_{i,\,i+1}\bigl(y_{i+1}-{y_{i}}{z_{i}}^{-1}z_{i+1}\bigr),
\label{02'}
\end{eqnarray}
where the following notations are made,  $\ddot{\xi}_i= \partial^2 \xi_i/\partial t^2$ and
$\xi=(x,y)$, in addition $"t"$ denotes the usual time. We will assume that near the nodes
spins are localized and quasi-periodic movements commit, $\xi_i(t)=\xi_i^0+\delta^\xi_i(t) $,
where  $\xi_i^0$ and $\delta_i(t)$ denote the  position of the equilibrium and
quasi-periodic function of the time respectively. Below we will study the statistical
properties of the system, which are formed on time scales $\tau>>\tau_0$, where
$\tau_0$ is a characteristic time of spins oscillation and obviously, in this
case; $\langle \ddot{x}\rangle_{\tau_0}=\langle \ddot{y}\rangle_{\tau_0}\approx 0$.
Averaging equations (\ref{02'}) on the period $\tau_0$ can be found:
\begin{eqnarray}
J_{i-1,\,i}\bigl(x_{i-1}-{x_i}{z_i}^{-1}z_{i-1}\bigr)+
J_{i,\,i+1}\bigl(x_{i+1}-{x_{i}}{z_{i}}^{-1}z_{i+1}\bigr)=0,
 \nonumber\\
J_{i-1,\,i}\bigl(y_{i-1}\,-{y_i}{z_{i}}^{-1}z_{i-1}\bigr)+
J_{i,\,i+1}\bigl(y_{i+1}-{y_{i}}{z_{i}}^{-1}z_{i+1}\bigr)=0,
 \label{03}
\end{eqnarray}
where for simplicity in equations the index $"^0"$ over of variables are omitted, i.e $x_i^0\rightarrow x_i,\,
y_i^0\rightarrow y_i$ and $z_i^0\rightarrow z_i$. As it is easy to verify these equations define the
 condition at which  the Hamiltonian (\ref{01}) in the $i$-\emph{th} node takes extremal value.

Solving the system of equations (\ref{03}), with respect to the variables $x_{i+1}$
and $y_{i+1}$, it can be found:
\begin{eqnarray}
x_{i+1}={C_x}/{J_{i,\,i+1}}, \qquad y_{i+1}={C_y}/{J_{i,\,i+1}},
 \label{04}
\end{eqnarray}
where the following notations are made:
$$
C_{x(y)}=\frac{A_{x(y)}-B_{y(x)}\bigl(C\pm \sqrt{D}\bigr)}{1+B_x^2+B_y^2},\quad
A_\eta={\eta_i}{z_i}^{-1}z_{i-1}-\eta_{i-1},\qquad
B_\eta={\eta_i}{z_i}^{-1}q_{i+1},
$$
$${D}= \bigl(1+B_x^2+B^2_y-A_x^2-A_y^2-C^2\bigr)>0,\quad
C=A_xB_y-A_yB_x,\qquad \eta=(x,y).$$

Now, for the Hamiltonian (\ref{01}) we can formulate conditions of the local minimum.
It is obvious that $i$-\emph{th} spin is in the stable equilibrium, if in the stationary
point the following inequalities are satisfied:
\begin{eqnarray}
A_{x_ix_i}(\textbf{\emph{s}}_i^0)>0,\qquad
A_{x_ix_i}(\textbf{\emph{s}}_i^0)A_{y_iy_i}(\textbf{\emph{s}}_i^0)-A^2_{x_iy_i}(\textbf{\emph{s}}_i^0)>0,
 \label{05}
\end{eqnarray}
where  $ A_{\eta_i\eta_i}={\partial^2 H}/{\partial \eta_i^2} $ and $
A_{x_iy_i}= {\partial^2 H}/{\partial x_i\partial y_i}; $ in addition
$\textbf{\emph{s}}_i^0$ denotes $i$-\emph{th} spin which is in a stable
equilibrium.\\
Using (\ref{02}), (\ref{03})  and  (\ref{05}), we can calculate the explicit forms of
the  second order derivatives:
\begin{eqnarray}
\hspace*{-1.5cm}
A_{\eta_i\eta_i}=(\eta^2_i+z^2_i)z_i^{-3}\Delta_i,\quad
A_{x_iy_i}=x_iy_iz_i^{-3}\Delta_i,\quad
\Delta_i=\bigl(J_{i-1,\,i}z_{i-1}+J_{i+1,\,i}z_{i+1}\bigr),
 \label{06}
\end{eqnarray}
and taking into account (\ref{05}) and (\ref{06}), it is easy to find the conditions
for a  local minimum of energy:
\begin{eqnarray}
A_{x_ix_i}=\bigl(1-y^2_i\bigr)z_i^{-3}\Delta_i>0,\quad
A_{x_ix_i}A_{y_iy_i}-A_{x_iy_i}^2=z_i^{-4}\Delta^2_i>0.
 \label{07}
\end{eqnarray}
Since by definition (\ref{02})  $z_i>0$, then both of the conditions in (\ref{07}) are satisfied:
  \begin{eqnarray}
\Delta_i=\bigl(J_{i-1,\,i}z_{i-1}+J_{i+1,\,i}z_{i+1}\bigr)>0.
 \label{08}
\end{eqnarray}
Thus, in each node the solutions defining the orientation of the spin in the state of the
local equilibrium  can be found, if we find such coupling constants $J_{i,\,i+1}$, for which
not only conditions (\ref{07}) or (\ref{08}) are satisfied, but also holds the inequality:
 \begin{eqnarray}
J_{i,\,i+1}^2\geq C_x^2+C_y^2>0.
 \label{09}
\end{eqnarray}
As  will be shown below, the additional condition (\ref{09}) will play an important role at simulation.

\section{Geometric properties of disordered $1D$ spin-chain}

\textbf{Theorem.} \emph{If the set of spatial spins;} $\{\emph{\textbf{s}}\}=
(\emph{\textbf{s}}_1,..,\emph{\textbf{s}}_n)$ \emph{forms the stable} 1$D$ \emph{spin-chain
 (see conditions (\ref{07})) then they necessarily are coplanar in the sense, that at
 parallel moving to the origin all spins lie in the same plane. }

\textbf{Proof.} Let us consider the three consecutive spatial spins
$\emph{\textbf{s}}_{i-1},\emph{\textbf{s}}_{i}$ and
$\emph{\textbf{s}}_{i+1}$ on the 1$D$ lattice. If we join the
origins of two consecutive  spins $\emph{\textbf{s}}_{i-1}$ and
$\emph{\textbf{s}}_i$, they will form a plane $\Lambda_0$. In this connection  arises the question
namely as subsequent spins are oriented relative to the plane $\Lambda_0$? Since these spins
are in the positions of local minima,  we can use the system of equations (\ref{03}) for defining
bonds between projections of three nearest-neighboring spins. In particular,
from the first equation in (\ref{03}) for the solution $z_{i+1}$, we can find the following expression:
\begin{eqnarray}
z_{i+1}=\frac{J_{i-1,\,i}\bigl(x_{i-1}z_i-x_iz_{i-1}\bigr)+J_{i,\,i+1}x_{i+1}z_i}{J_{i,\,i+1}x_i}.
 \label{10}
\end{eqnarray}
Substituting $z_{i+1}$ into the second equation in (\ref{03}),  the expression of bond between
projections of two spins $\emph{\textbf{s}}_{i-1}$ and $\emph{\textbf{s}}_i$ can be found:
\begin{eqnarray}
x_{i-1}y_{i}-x_{i}y_{i-1}=\frac{J_{i,\,i+1}}{J_{i-1,\,i}}\bigl(x_{i+1}y_i-x_iy_{i+1}\bigr).
 \label{11}
\end{eqnarray}
The spin $\emph{\textbf{s}}_{i+1}$ is a parallel to the plane $\Lambda_0$, if the
following equation is satisfied:
\begin{eqnarray}
\Biggl|\begin{array}{ccc}
  x_{i-1} &  y_{i-1}  & z_{i-1}  \\
  x_i & y_i & z_i \\
x_{i+1} & y_{i+1} & z_{i+1}
\end{array}
\Biggl|=0.
 \label{12}
\end{eqnarray}
We can write the equation (\ref{12}) in the explicit form:
$$
\det|\cdot|=x_{i-1}y_iz_{i+1}+x_{i+1} y_{i-1}
z_{i+1}+x_iy_{i+1}
z_{i-1}-x_{i+1}y_iz_{i-1}-x_iy_{i-1}z_{i+1}-x_{i-1}y_{i+1}z_i=
$$
$$
x_{i+1}\Bigl[-\frac{J_{i,\,i+1}}{J_{i-1,\,i}}\bigl(y_{i+1} z_i-y_i
z_{i+1}\bigr)\Bigr]+y_{i+1}\Bigl[\frac{J_{i,\,i+1}}{J_{i-1\,,i}}\bigl(x_{i+1}
z_{i}-x_{i} z_{i+1}\bigr)\Bigr]
+z_{i+1}\Bigl[-\frac{J_{i,\,i+1}}{J_{i-1,\,i}}\bigl(x_{i+1}
y_{i}-x_iy_{i+1}\bigr)\Bigr],
$$
Finally, using the expression (\ref{11}) it is easy to show that:
$$
\det|\cdot|=\frac{J_{i,i+1}}{J_{i-1,i}}\Bigl\{x_{i+1}\bigl(
y_i z_{i+1}- z_iy_{i+1}-y_i z_{i+1}+z_i y_{i+1}\bigr)+x_i\bigl(
y_{i+1} z_{i+1} -y_{i+1} z_{i+1}\bigr) \Bigr\}=0.
$$
\emph{Thus the theorem is proved. }

Note that, the specified geometric property allows simplifying  the Hamiltonian (\ref{01}).\\
Let us consider  the  set of spins in the spherical coordinate system $(\alpha_i,\theta_i,\vartheta_i)$.
In the new coordinates for two consecutive spins, we can write the following relationship:
\begin{eqnarray}
\emph{\textbf{s}}_i\emph{\textbf{s}}_{i+1}=||\emph{\textbf{s}}_i||\cdot||\emph{\textbf{s}}_{i+1}||
=\cos(\alpha_i-\alpha_{i+1}),
 \label{13}
\end{eqnarray}
where $(\alpha_i,\alpha_{i+1})\in[-\pi,+\pi]$  are angles of the respective spins in planes parallel
to plane $ \Lambda_0$.\\
Using (\ref{13}) Hamiltonian (\ref{01}) can be written as:
\begin{eqnarray}
H=-P(\theta,\vartheta)\sum_{i=1}^n
J_{i,i+1}\cos(\alpha_i-\alpha_{i+1}),
 \label{14}
\end{eqnarray}
where, as it follows from the proof of the proposition,
$\theta=\theta_1=...=\theta_n\in (-\pi,+\pi]$ and
$\vartheta=\vartheta_1=...=\vartheta_n\in[0,\pi]$. In addition, the pair of angles $ (\Theta, \vartheta) $
determines the orientation of the plane $\Lambda_0$ in 3$D$ space. It is natural to propose that
$P(\theta,\vartheta)$ is a homogeneous distribution function from
angles, which is normalized on unit $\int\int P(\theta,\vartheta) d\theta
d\vartheta =1$.
\begin{figure}
\center
\includegraphics[height=55mm,width=105mm]{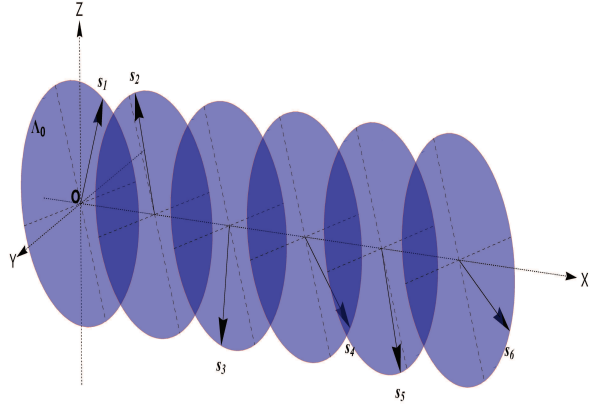}
 \caption{\emph{The disordered $1D$ spin chain where spins lie in planes  parallel to the plane} $\Lambda_0$.}
 \label{Fig1}
\end{figure}

For finding the extreme value of the  Hamiltonian (\ref{14}) in nodes, let us consider
the first derivative by the angle $\alpha_{i}$:
\begin{eqnarray}
\frac{dH}{d\alpha_i}=P(\theta,\vartheta)\bigl[J_{i-1,\,i}\sin(\alpha_{i-1}-\alpha_{i})-
J_{i,\,i+1} \sin(\alpha_{i}-\alpha_{i+1})\bigr].
 \label{15}
\end{eqnarray}
It is obvious that derivatives of  Hamiltonian (\ref{14}) by angles $\theta$ and $\vartheta$
are identically equal to zero. Now by equating the expression (\ref{15}) to zero and solving
it, we will get two possible solutions for a stationary point:
\begin{eqnarray}
\alpha_{i+1}=\alpha_{i}-\arcsin
\biggl[\frac{J_{i-1,\,i}}{J_{i,\,i+1}}\sin(\alpha_{i-1}-\alpha_{i})\biggr],\\
 \alpha_{i+1}=\alpha_{i}+\pi+\arcsin
\biggl[\frac{J_{i-1,\,i}}{J_{i,\,i+1}}\sin(\alpha_{i-1}-\alpha_{i})\biggr]. \nonumber
 \label{16}
\end{eqnarray}
The condition on existence of these solutions in the region of  real numbers is equivalent
to the following inequality:
\begin{eqnarray}
\hspace*{-1cm}
-1\leq
\frac{J_{i-1,\,i}\sin(\alpha_{i-1}-\alpha_{i})}{J_{i,\,i+1}}\leq1,\quad
or\quad |J_{i,\,i+1}|\geq |J_{i-1,\,i}\sin(\alpha_{i-1}-\alpha_{i})|.
 \label{17}
\end{eqnarray}
Using two equations from  (\ref{16}) and  substituting  $ i$  instead of $i-1$ we can
find the value of $J_{i-1,i} \sin(\alpha_{i-1}-\alpha_i )$, and for  both solutions result will be same:
$$
J_{i-1,\,i}
\sin(\alpha_{i-1}-\alpha_i)=J_{i-1,\,i}\frac{J_{i-2,\,i-1}\sin(\alpha_{i-2}-\alpha_{i-1})}{J_{i-1,\,i}}
=J_{i-2,\,i-1}\sin(\alpha_{i-2}-\alpha_{i-1}).
$$
It is clear that by continuing this process we will get:
\begin{eqnarray}
J_{i-1,\,i}
\sin(\alpha_{i-1}-\alpha_i)=J_{1,2}\sin(\alpha_{1}-\alpha_2).
 \label{18}
\end{eqnarray}
Using (\ref{18}) we can transform condition (\ref{17}) to
\begin{eqnarray}
|J_{i,\,i+1}|\geq |J_{1,2}\sin(\alpha_{1}-\alpha_2)|.
 \label{19}
\end{eqnarray}
Let us  note that the angles, $\alpha_1,\alpha_2$ and also the coupling constant,
$J_{1,2}$ in  condition  (\ref{18}) as an initial conditions of problem are specified.
Finally we can write the condition of the local minimum energy in the arbitrary
 $i$-\emph{th} node:
$$
\frac{\partial^2
H}{\partial\alpha_i^2}=P(\theta,\vartheta)\bigl[
J_{i-1,\,i}\cos(\alpha_{i-1}-\alpha_i)-J_{i,\,i+1}\cos(\alpha_{i}-\alpha_{i+1})\bigr]>0.
$$


\section{The statistical ensemble of  1$D$ disordered spin-chains}
As it is easy to verify solutions of equations (\ref{04}) satisfying the
inequalities (\ref{07}) can be of two types:

a. If $J_{i-1,\,i}\emph{\textbf{s}}_{i-1}\cdot\emph{\textbf{s}}_i \leq
0$ and $|J_{i,\,i+1}|>|J_{i-1,\,i}|$, then there is only  one solution,
which we denote by;  $\emph{\textbf{s}}_{i+1}^+$ (\emph{queen}), and
respectively,

b. If $J_{i-1,\,i}\emph{\textbf{s}}_{i-1}\cdot\emph{\textbf{s}}_i > 0$
and $|J_{i,\,i+1}|\geq|J_{0,1}|\cdot
|\emph{\textbf{s}}_0\times\emph{\textbf{s}}_1|$, then
$\emph{\textbf{s}}_{i+1}^+$ is the solution, in addition there is
another  solution; $\emph{\textbf{s}}_{i+1}^-$ (\emph{drone}) under
the condition that, $|J_{i,\,i+1}|<|J_{i-1,\,i}|$.

Recall,  that the solutions which are denoted with signs $^{"+"}$ and $^{"-"}$ are
characterized as follows, if the previous solution is the \emph{queen}  $^{"+"}$,
then it is possible to find  up two different solutions $\emph{\textbf{s}}_{i+1}^+$ and
$\emph{\textbf{s}}_{i+1}^-$, while after the  \emph{drone}  $^{"-"}$ the solution only
one $\emph{\textbf{s}}_{i+1}^+$. Taking into account this we
can construct solutions graphically in the form of separate
\emph{Fibonacci subtrees} ($\widehat{FsT}_i$) (see Fig. 2).

The mathematical expectation of the branching   depending on the height of
$\widehat{FsT}_i$ can be calculated by the following formula:
\begin{equation}
\hspace*{-1cm}
M(n)=M(n-1)\bigl\lfloor\,{(2\xi_n)}\bigr\rfloor=
\bigl\lfloor2^{n\eta(n)}\bigr\rfloor, \quad \eta(n)=
 1+n^{-1}\sum_{k=1}^n\log_2(\xi_k)>0,\label{21}
\end{equation}
where $M(n-1)$ the number of  the branching  at the height $(n-1)$, while $\xi_k$ denotes
a random coefficient which belongs to the interval $[1/2,1]$. Note, that for
simplification of the formula  (\ref{21}) designating the subtree's number $i$ is omitted.
Since, an each $\widehat{FsT}_i$  consists of the set of nodes and the set of edges
(the set of  constants $\{J\}=[J_{1,2},J_{2,3},...J_{n-1,n}]$ therefore it can
be represented as a graph $G_i(n)\cong\{g_j(n),j\in M\}$, where $g_j(n)$ denotes a
random string by length $n$ which is characterized by Kolmogorov's complexity \cite{Kolm,Li}.
\begin{figure}
\center
\includegraphics[height=185mm,width=155mm]{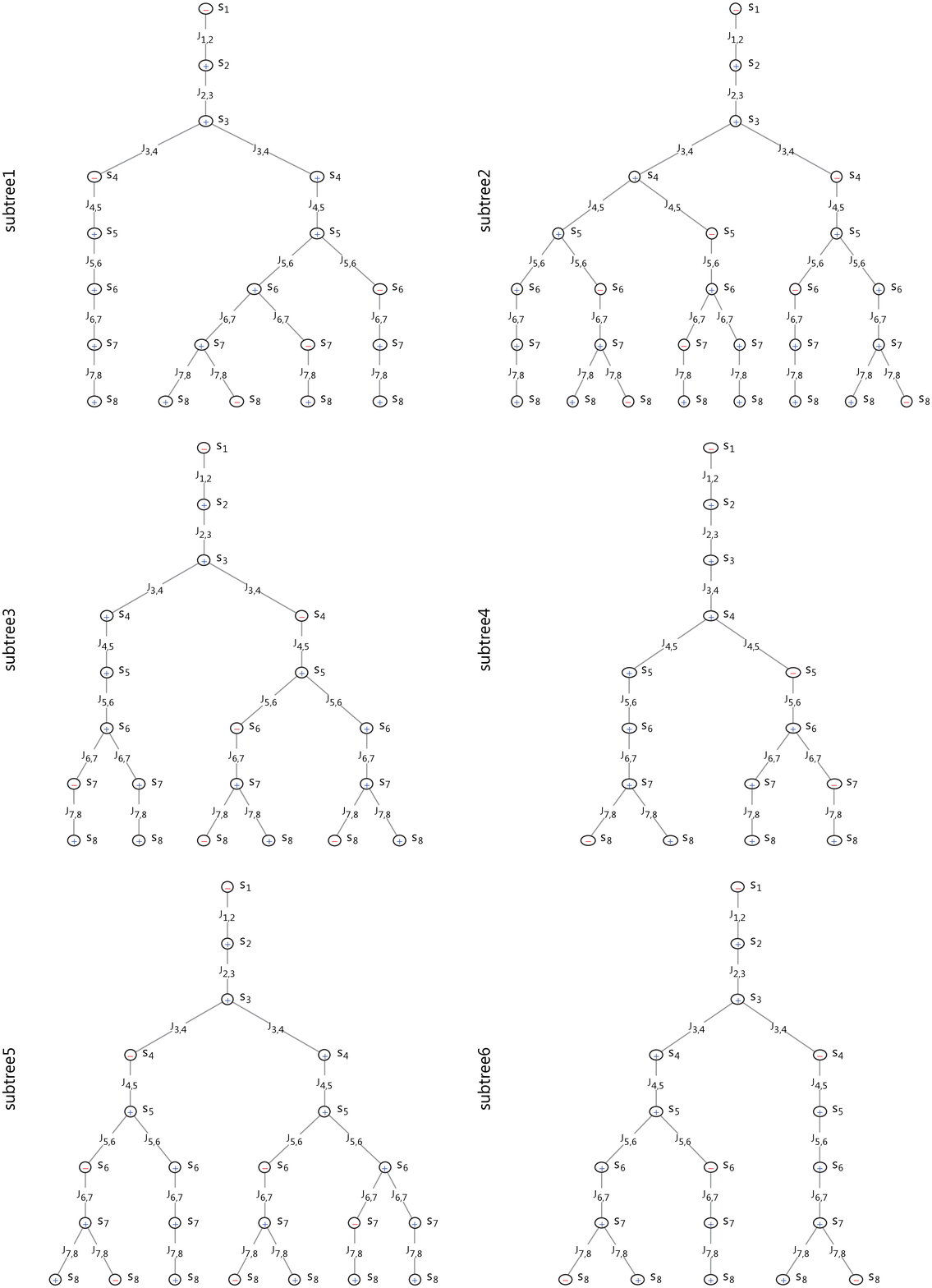}
\caption{\emph{The six different Fibonacci subtrees (graphs)  with the  height  8. All these graphs
are growing from the same initial data} (\emph{root}) \emph{in result of the six independent numerical
experiments. Note, that the same symbols }$s_i$ \emph{and} $J_{i,j}$ \emph{on different graphs can have
completely  different values.}}
 \label{Fig2}
\end{figure}

Note that each Fibonacci subtree (graph) depending on its height $n$ can be represented
itself as a random process. For their comparing we can formulate the following criterion:

\textbf{Definition.} \emph{Two graphs with the same height; $G_1(n)$ and $G_2(n)$ are
equivalent with a given accuracy $O(\epsilon)$, where $\epsilon\ll1$, if the following conditions take place:}

1) \emph{The difference of Shannon's entropy of the two Fibonacci subtrees (graphs)
satisfies:}
\begin{equation}
\bigl|S^1(n)- S^2(n) \bigr|\leq \epsilon,\qquad
S^{1(2)}(n)=-\sum_{i=1}^{n}M_i^{1(2)}\ln M_i^{1(2)},
 \label{22}
 \end{equation}
 \emph{where}  $S^1(n)$ \emph{and} $S^2(n)$ \emph{denote the Shannon's
entropies of graphs $G_1(n)$ and $G_2(n)$, in addition}
$M_i^{1}$ \emph{and} $M_i^{2}$  \emph{are the branching numbers of corresponding graphs
on the}  $i$-\emph{th height,}

 2)  \emph{the difference of average polarizations of two graphs in per one spin
 satisfies:}
 \begin{equation}
 \Bigl\|{{\frac{1}{n}\sum_{i=1}^{n}\Bigl(
 \emph{\textbf{s}}_i^{(1)}-
\emph{\textbf{s}}_i^{(2)}\Bigr)}}\Bigr\|\leq\epsilon,
 \label{23}
\end{equation}
\emph{where} $\emph{\textbf{s}}_i^{(1,2)}=\sum_{G_{1,2}(i)}\emph{\textbf{s}}_{i}$
\emph{denotes the total value of spins on the corresponding graph at the $i$-th height,}

3) \emph{the difference of the average energies of two graphs in per one
spin satisfies:}
\begin{equation}
\frac{1}{n}\sum_{i=1}^n\Bigl|\frac{1}{m_1}{{\sum_{j=1}^{m_1}
J^{(1)}_{i,\,i+1; (j)}
\emph{\textbf{s}}_{i;(j)}\emph{\textbf{s}}_{i+1;(j)}-\frac{1}{m_2}\sum_{j=1}^{m_2}
J^{(2)}_{i,\,i+1; (j)}
\emph{\textbf{s}}_{i;(j)}\emph{\textbf{s}}_{i+1;(j)}}}\Bigr|\leq\epsilon,
\label{24}
\end{equation}
where $ m_1=M^1_n$ and $m_2=M^2_n$.

In the case when at least one condition from (\ref{22})-(\ref{24}) is violated,
we will consider that $G_1(n)$ and $G_2(n)$ are inequivalent or independent.

Thus, for calculations of different physical parameters of the statistical ensemble, it is
necessary to take into account the contribution of all independent graphs (set of graphs)
$\{G(n)\}_N=[G_1(n), ...G_i(n),...]$, where $i=\overline{1,N}$ and $N$ the number of graphs.

As mentioned above, the system of equations (\ref{04}) which satisfies conditions
(\ref{07})-(\ref{08}) in each node can have up to two solutions. The latter  means that the
number of solutions on the step $n$ due to branching will be of order $M(n)\propto 2^n$ and
correspondingly, the calculation problem of the statistics even of one graph algorithmically
is a $NP$  hard problem, since at increasing of spins number the number of solutions grows exponential.

Evaluation of the computational complexity of the statistics for a single graph gives:
\begin{equation}
 K_t(n)\propto M(n)K_s(n),
 \label{25}
\end{equation}
where  $K_s(n)$ denotes the Kolmogorov complexity of the string $g_j(n)$, while
$K_t(n)$ denote the complexity of the graph $G_i(n)\subset\{G(n)\}_N$. The computational
complexity of the ensemble, which is represented as the set $\{G(n)\}_N$, obviously will
be order; $K_{ens}\propto NM(n)K_s(n)$.

The mathematical expectation of random variable $f$ characterizing the
ensemble $\{G(n)\}_N$ can be calculated by the formula:
\begin{equation}
E[f]=\bar{f}=\frac{\sum_{i=1}^{N} w_i\bar{f}_i}{\sum_{i=1}^{N} w_i}, \qquad w_{i}=N_i/\bar{N},
\label{26}
\end{equation}
where $N_i$ and $\bar{N}$ denote the number of strings of the graph $G_i(n)$ and the
total number of strings in the ensemble respectively, in addition $\bar{f}_i=\sum_{G_i(n)}f$
denotes the expectation of a random variable  $f$ on the $G_i(n)$, which is calculated similarly
to formula (\ref{26}).

From the point of view of statistics, it is important to investigate the ensemble in
the state of the statistical equilibrium. This as a rule is realized at $N>>1$ and when the
average value of random variable $f$ almost surely converges to the expected value
\cite{Grimmett}:
$$
{\mathrm{Pr}\Bigl(\lim_{N\to\,\infty}
\bar{f}_N=\bar{f}\Bigr)}=1,
$$
where $f_1,f_2,...$ are infinite sequence of  Lebesgue integrable
random variables with the expected values $E[f_1]=E[f_2]=...=\bar{f}.$

\textbf{Lemma.}\emph{ If statistical weights of all independent graphs $G_i(n)\subset \{G(n)\}_N$
are approximately the same it can be shown that the statistical weights of all strings
$g_j(n)\subset \{G(n)\}_N$ are equal exactly. In this case we can use  the law of large numbers and
simplify the expression (\ref{26}) writing it as:}
\begin{equation}
E[f]=\bar{f}=\frac{1}{N}\sum_{j=1}^N\tilde{f}_j+O({N^{-1/3}}),
\label{27}
\end{equation}
\emph{where $\tilde{f}_j=\sum_{g_j}f$ denotes the expectation of the random
variable $f$ on a randomly selected string} $g_j(n)\subset G_i(n)$.

Note that the asymptotic convergence to the limit
value in the expression (\ref{27}) occurs with accuracy $\propto {N^{-1/3}}$ due to
the fact that the spins are three-dimensional.

Thus, the computation of statistical parameters of the disordered spin system by
the formula (\ref{26}) is algorithmically equivalent to solving of
$\mathbb{NP}$ hard problem (the left scheme in Fig. 3).  In the case when the
ensemble is in the state of statistical equilibrium then the numerical simulation can be realized by
the formula (\ref{27}) and respectively by the algorithm $\mathbb{P}$ (the right scheme in Fig. 3)
having the polynomial complexity.
\begin{figure}
\center
\includegraphics[height=80mm,width=80mm]{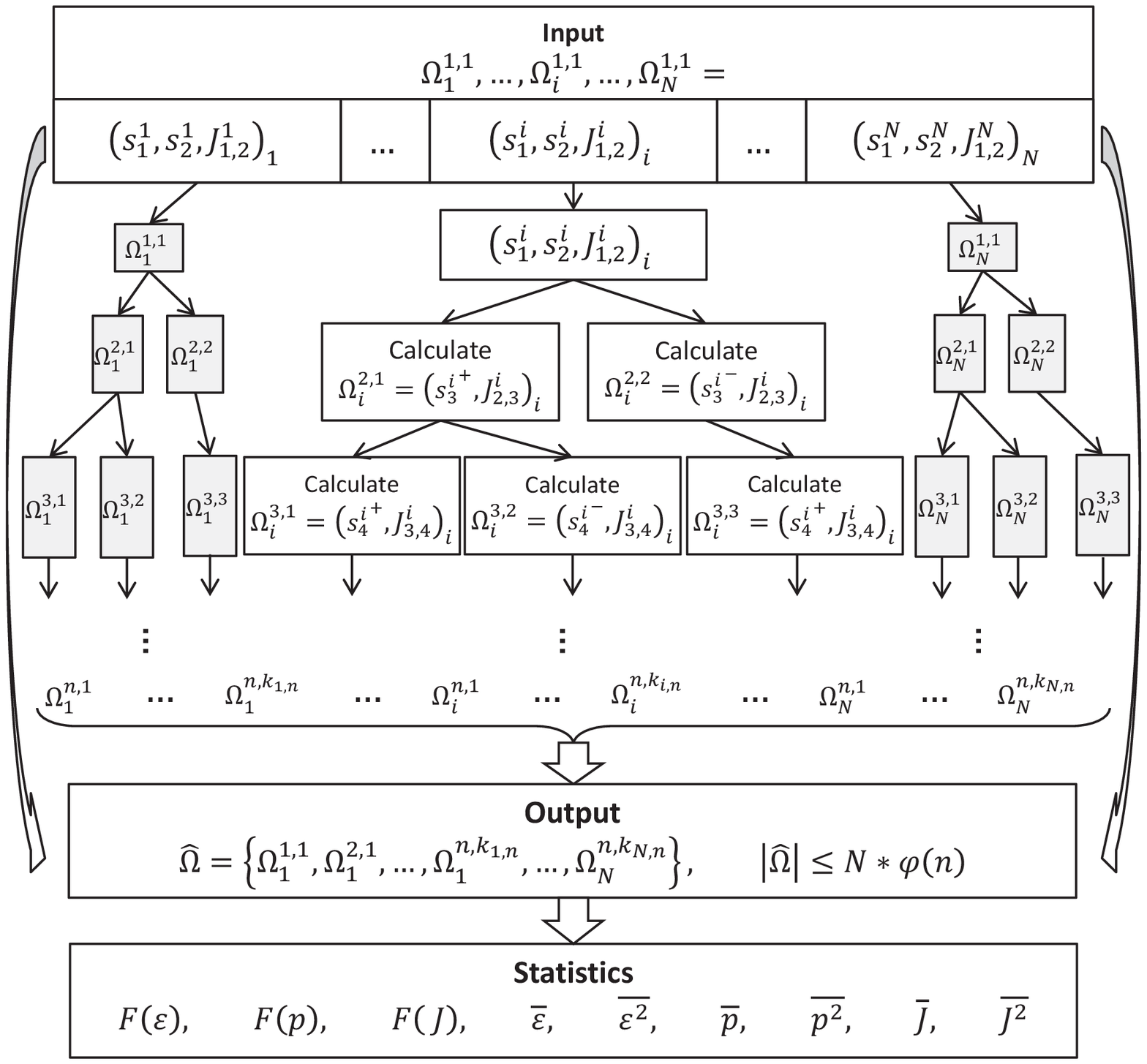}
\includegraphics[height=80mm,width=80mm]{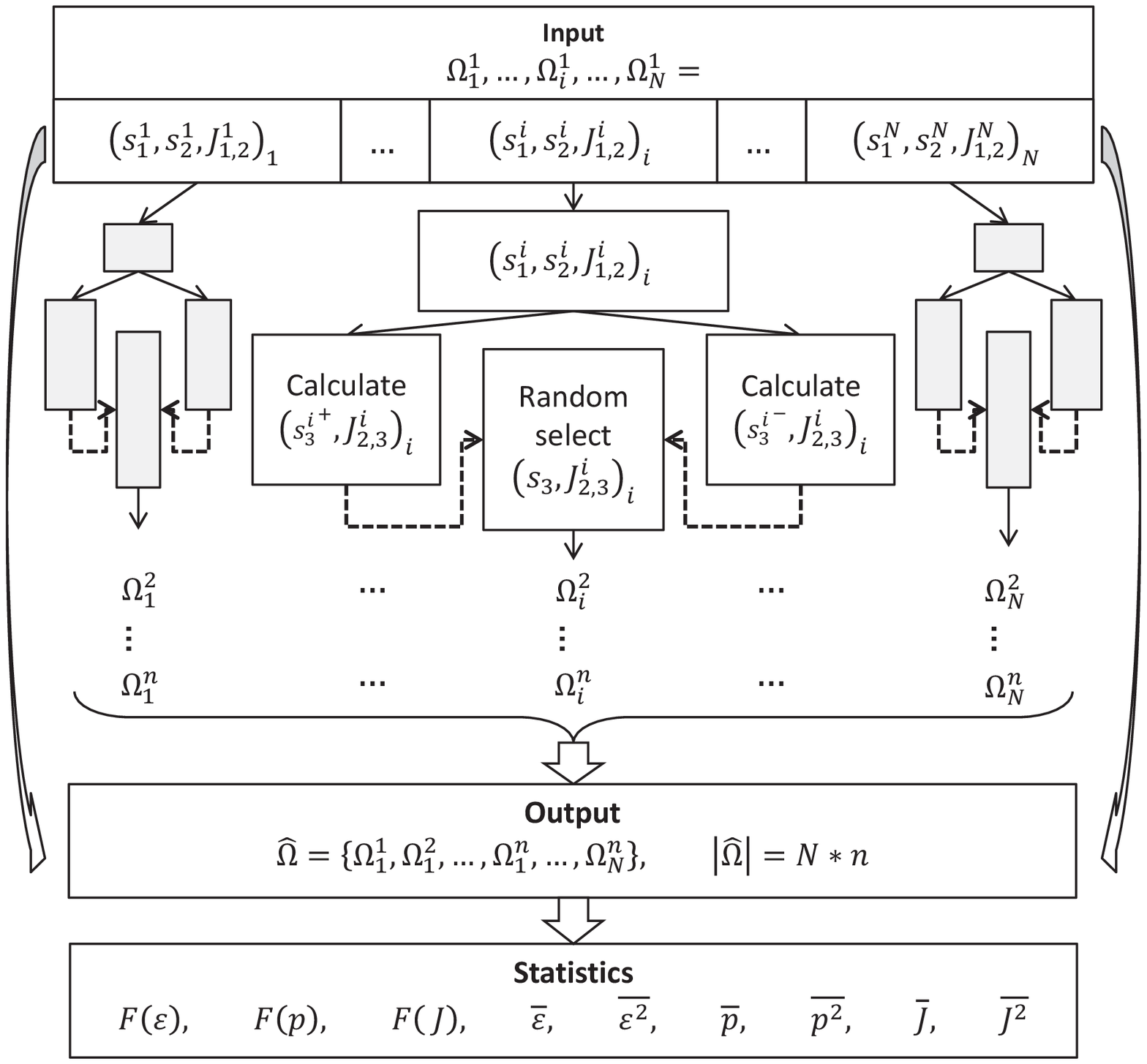}
\caption{\emph{The left scheme describes algorithm, which allows to sort out all graphs of the
ensemble and, correspondingly,  this algorithm in the future  will be called $\mathbb{NP}$
algorithm, while the right one describes} $\mathbb{P}$ \emph{algorithm which allows implementing
calculations of the problem in a polynomial time.}}
 \label{Fig3}
\end{figure}


\section{The numerical experiments}
As it has been said, usually the problems of spin glasses are studied in the framework
of the partition function representation by using Monte Carlo simulation methods, which
however does not allow to answer on many important questions of the statistical ensemble.
In particular, is an important problem is that the spin glass in the state of a statistical
equilibrium generally speaking is in a metastable state and has some distribution near the
\emph{ground state}, while Monte Carlo simulation methods are adapted for calculations only
the \emph{ground state}. It is clear, that the influence and contribution of this distribution
on different properties and values of parameters of a spin glass may be accounted, if the
numerical simulation of the spin system to spend from first principles of classical mechanics.
\begin{figure}
\includegraphics[height=55mm,width=70mm]{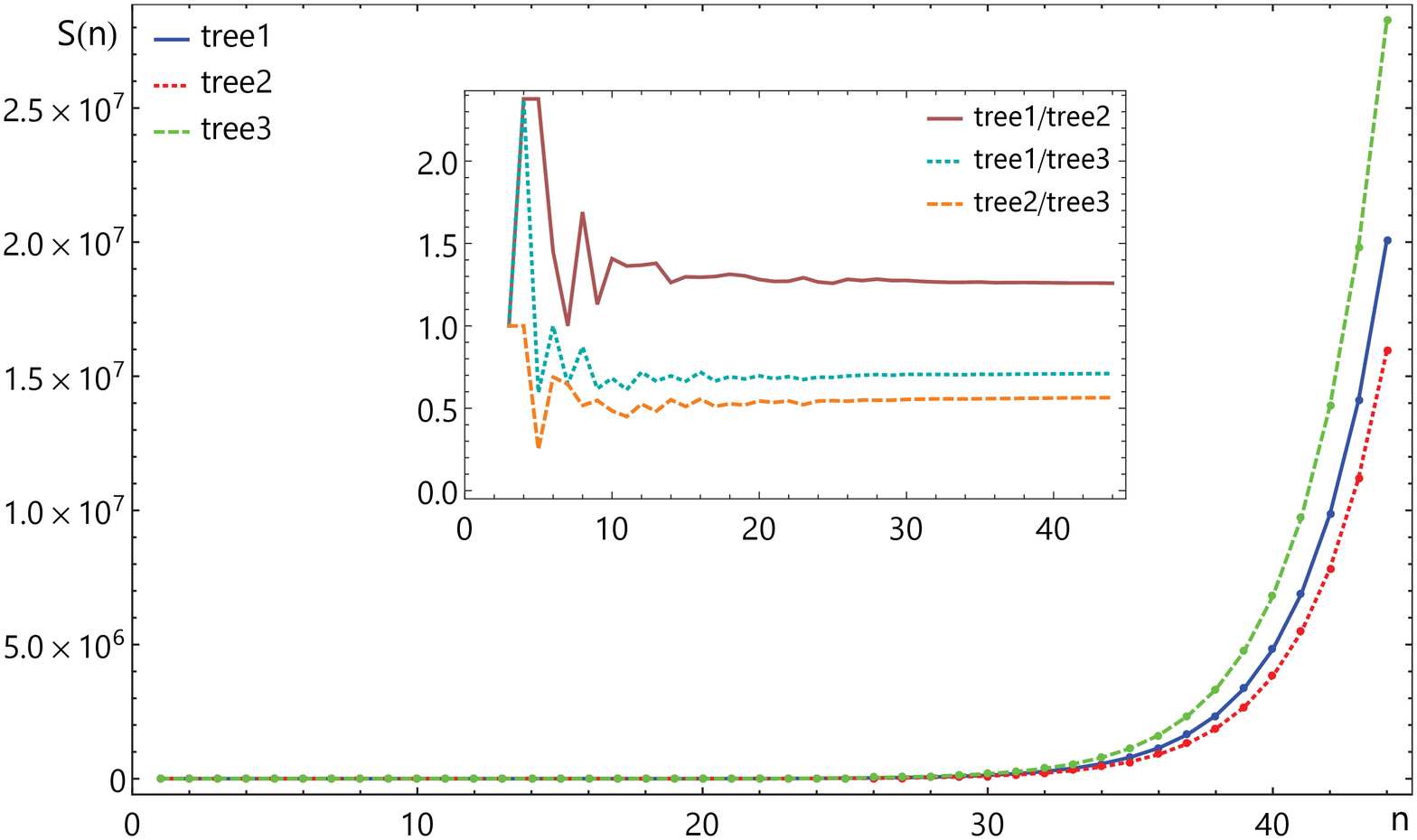}
\qquad
\includegraphics[height=55mm,width=70mm]{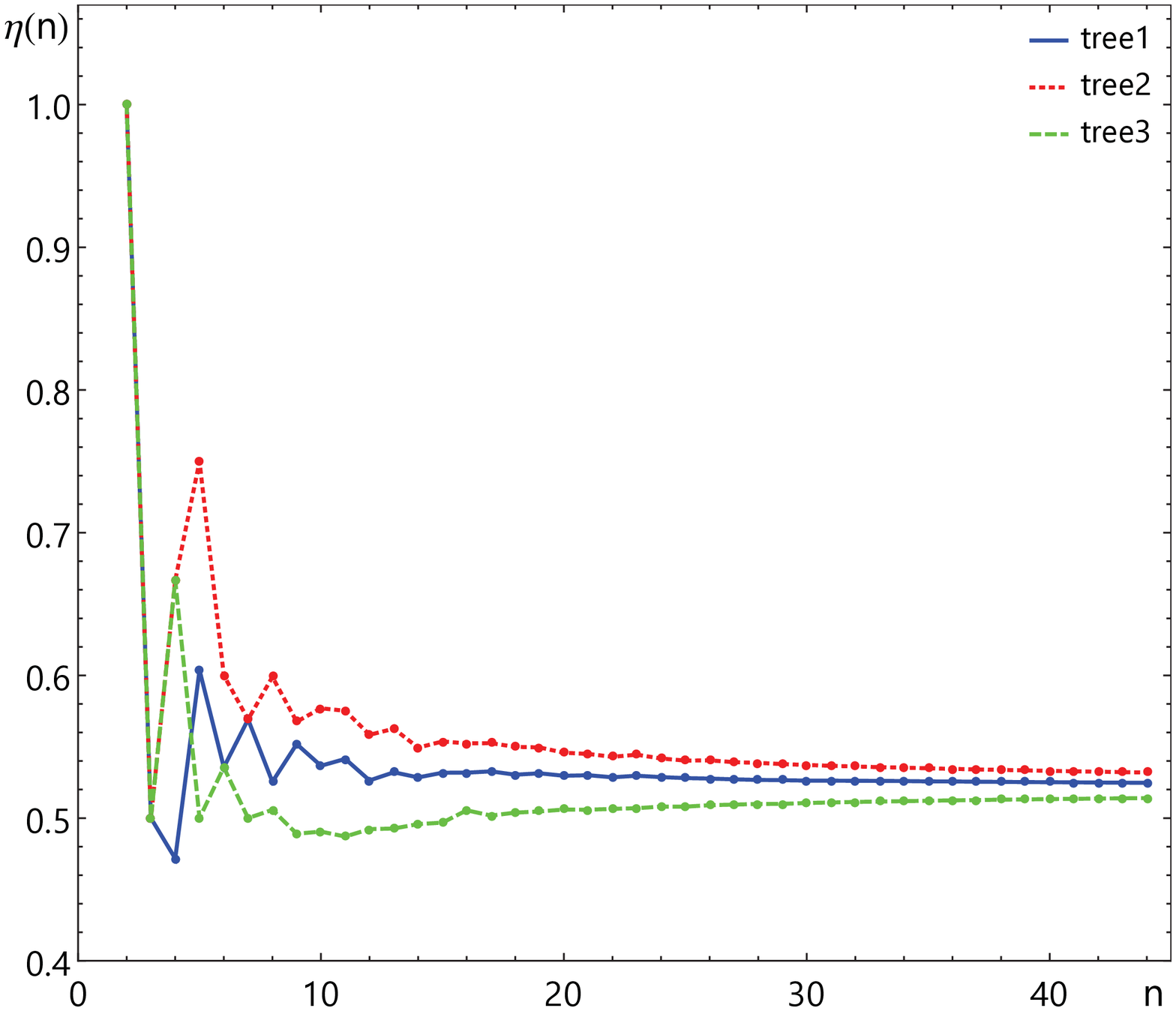}
\caption{\emph{In the left figure are shown the entropies of graphs (subtrees) depending on their height
(the red, blue, and green lines), while in the small frame are shown curves of relations
of corresponding entropies. In the right figure are shown curves of the branching factor $\eta(n)$ of
different graphs depending on their height.}}
\end{figure}

\textbf{Hypothesis.} \emph{If the Heisenberg $1D$ spin glass is in the state of the
statistical equilibrium, then the computational $\mathbb{NP}$ hard problem with the prescribed
accuracy $\epsilon$ can be reduced to the $\mathbb{P}$ problem.}

It should be noted that at  performing of numerical simulations with the same initial data,
everytime  $"t"$ we find a new set of graphs $\{G(n)\}_N^t$ (see Fig. 2), nevertheless we expect
that  in the limit of statistical equilibrium all these sets must be identical in terms of
statistical properties $\{G(n)\}_N^{t_0}\propto\{G(n)\}_N^{t_1}...\propto\{G(n)\}_N^{t_n}$ and  this is
the assumption of the hypothesis.
It is obvious, if we  prove that all strings  in the statistical ensemble $\{G(n)\}_N$, have the
equal weight then this allows to use the law of big numbers  and  to reduce
$\mathbb{NP}$ hard problem to $\mathbb{P}$ problem with the prescribed accuracy.
\begin{figure}
\includegraphics[height=55mm,width=70mm]{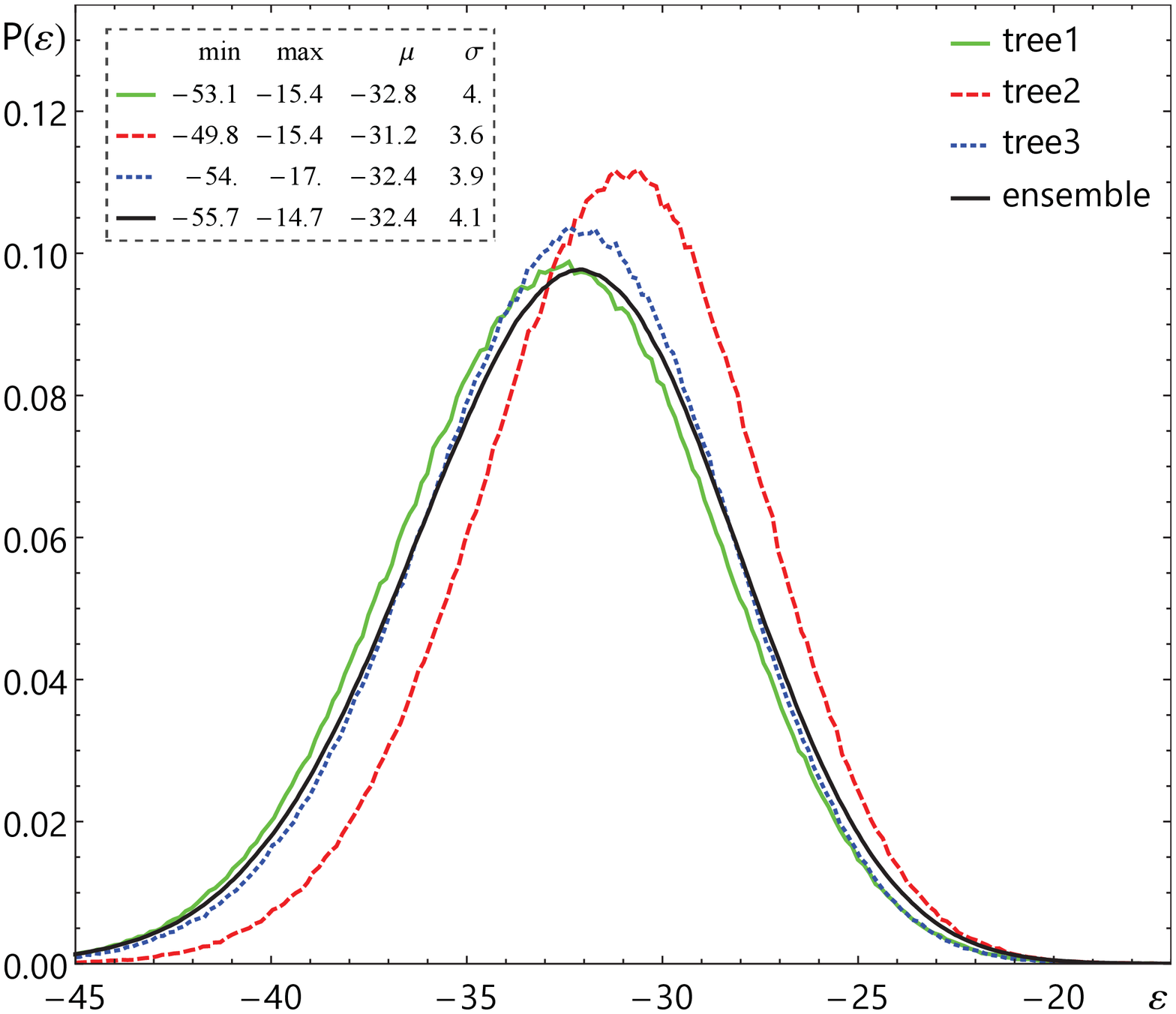}
\qquad
\includegraphics[height=55mm,width=70mm]{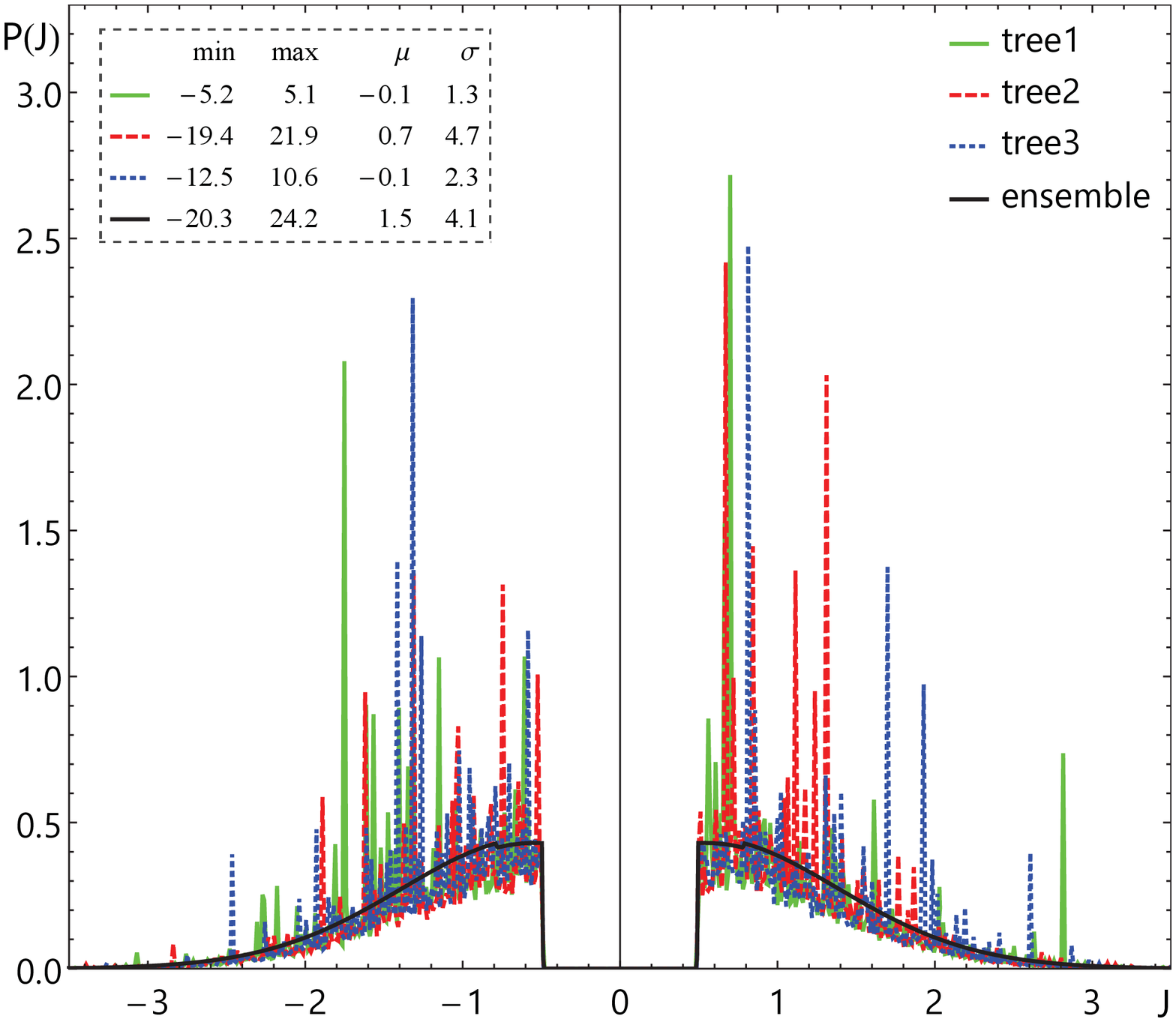}
\caption{\emph{On the left picture are shown the distributions of strings' energies of
the length} $45$ \emph{in the three different graphs (red, green and blue lines) which
\emph{grow} from the same root and correspondingly the black curve, which
shows the energy distribution in statistically equilibrium ensemble $\{G(n)\}_N$,
where all graphs from  one root  are growing. In addition, in tables adduced
the important parameters that characterize the corresponding distributions; the maximal
and minimal values, the average value of parameter $\mu=\int xP(x)dx$ and the dispersion
$\sigma $. Note that the simulation has been conducted by $\mathbb{NP}$ algorithm (see the left
 scheme  on Fig. 3). }}
\end{figure}

For a detailed study of the properties of graphs  and their contributions to the statistics
of the ensemble, we will consider two possible cases; when graphs are growing from one single
\emph{root} and, respectively, when they grow from different  \emph{roots}.

At first let us consider one set of initial data $\Omega_i^1$ (root) which includes orientations of
the first two spins of the chain and the coupling constant between them which  are generated randomly
from the corresponding homogeneous distributions. Using  the system of recurrence equations (\ref{04}), with
consideration of inequality conditions (\ref{07}), we  perform successive calculations of spin-chain.
Recall that this system of equations connects three consecutive spins, so that knowing the configuration
of two previous spins, we can generate from lognormal distribution \cite{lognormal} a random constant
$J_{i,\,i+1}$ and exactly to calculate the orientation of the spin in the subsequent node.
Conducting the consecutive node-by-node calculations on the $n$-\emph{th} step, we generate a random
graph $G_i(n)\subset\{G(n)\}_N$ at internal nodes of which spins are in local minima of energies.
With regard to the spins in the external nodes, it is assumed that they satisfy the conditions of
local minima of energy, on the basis of other considerations.

The simulation using the $\mathbb{NP}$ algorithm shows, that all three graphs which \emph{grow} from
one \emph{root} are independent, by the  criteria (\ref{22})-(\ref{24}). In particular
the numerical simulations show that depending on  height of the graph,  the Shannon's entropy
\emph{grows} an exponential, in all cases starting with $n\simeq15$ (see the left picture on Fig. 4).
The ratios of entropies, as shown in  Fig. 4, for the $n>15$  take values $\propto O(1)$ that means
in the ensemble $\{G(n)\}_N$ the weights of separate graphs are approximately equal. The
weight of individual branches in the statistical ensemble obviously will be the inverse of weights of
graphs to which they belong. In other words all branches in the ensemble have the same weight. Note
that the same picture is observed when graphs are \emph{growing} from different \emph{roots}. In
this case all graphs  are also independent and the parameter of branching, at increasing
of string length as in the previous case converges to the value $\eta(n)=0.55$ (see the right picture
on Fig. 4). When the length of string $n<15$ then in the behaviour of entropy an oscillating
character is observed  (see Fig. 4), that is characteristic of the discrete systems and manifests
itself as a \emph{size effects}. We carried calculations of distributions of different parameters
on the example of three  graphs and also of the ensemble of graphs which \emph{grow} from the same
\emph{root}. As the calculations show, the energies distributions for three graphs and the ensemble,
$\{G(n)\}_N$ by criterion of Kullback-Leibler distance are close enough \cite{Kul-Lei}, while
distributions of the spin-spin coupling are sufficiently far, by the same criterion (see Fig. 5).

So, we have shown that there are necessary and sufficient conditions for performing of the \textbf{lemma}.\\
\begin{figure}
\includegraphics[height=65mm,width=70mm]{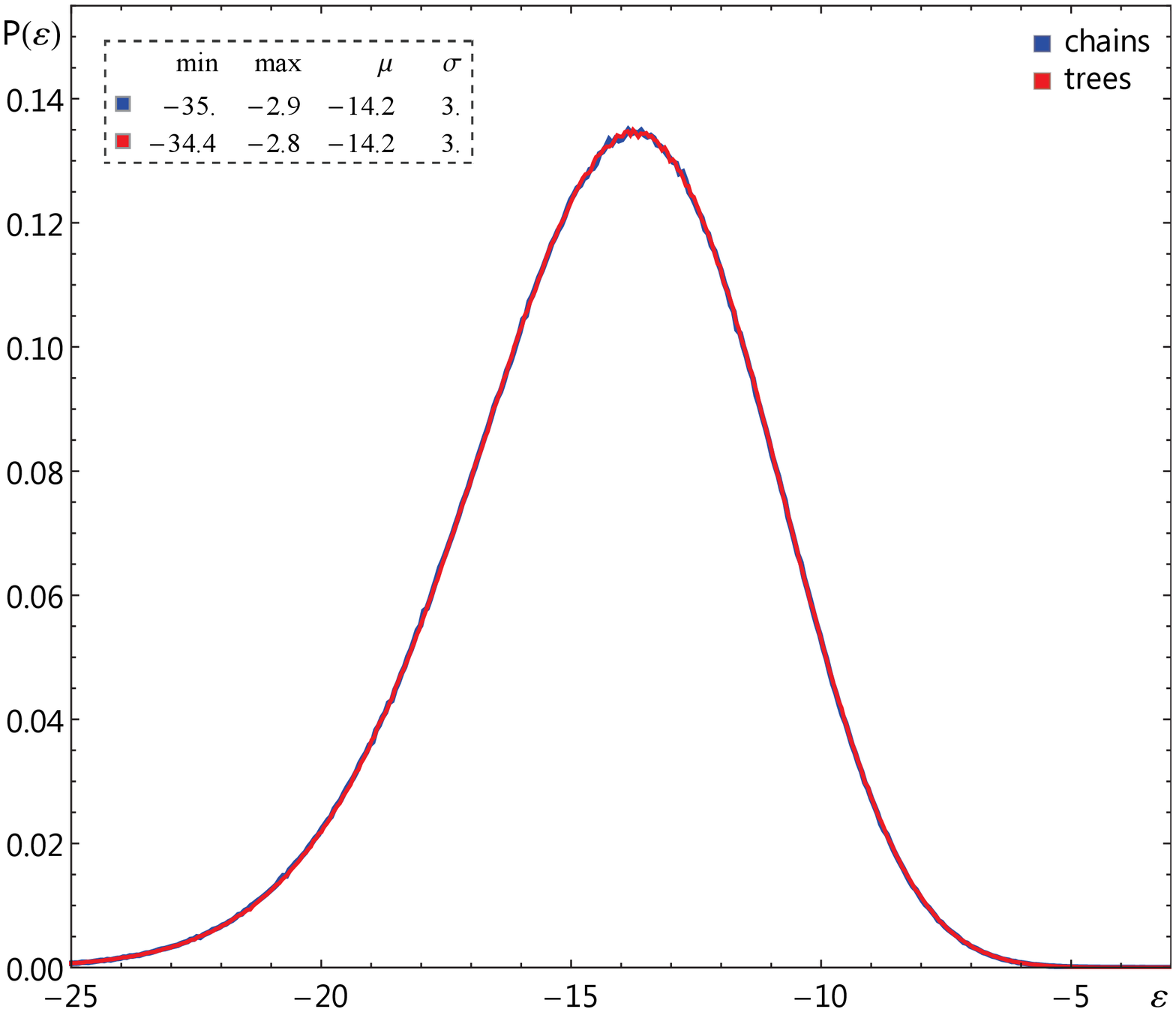}
\qquad
\includegraphics[height=65mm,width=70mm]{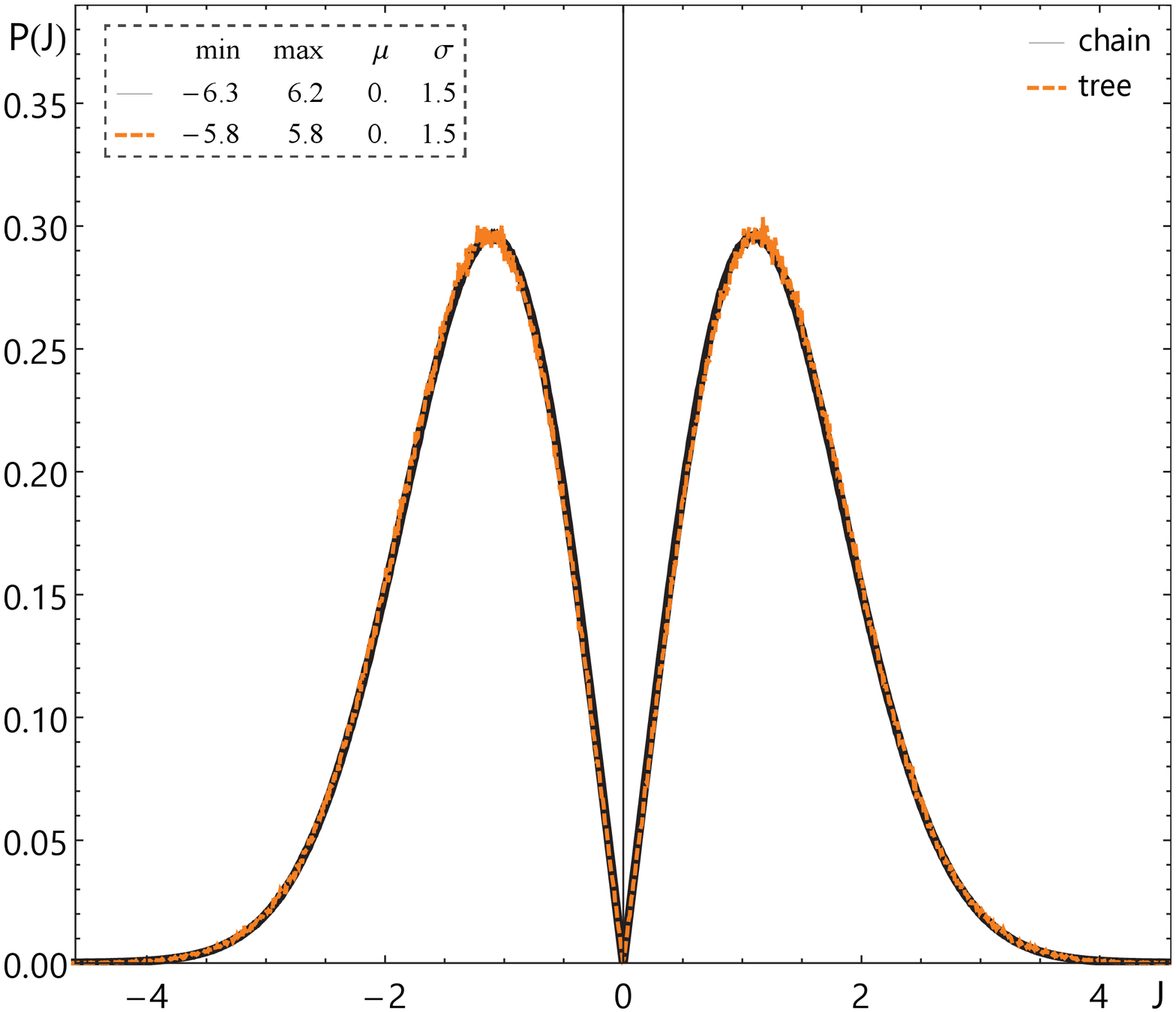}
\caption{\emph{The distributions of energies and spin-spin coupling constant.
 The black curves denote the results of calculations using $\mathbb{P}$ algorithm, while
 beige curves are  constructed  in result of
calculations by $\mathbb{NP}$ algorithm.}}
\end{figure}
\begin{figure}
\qquad \includegraphics[height=65mm,width=130mm]{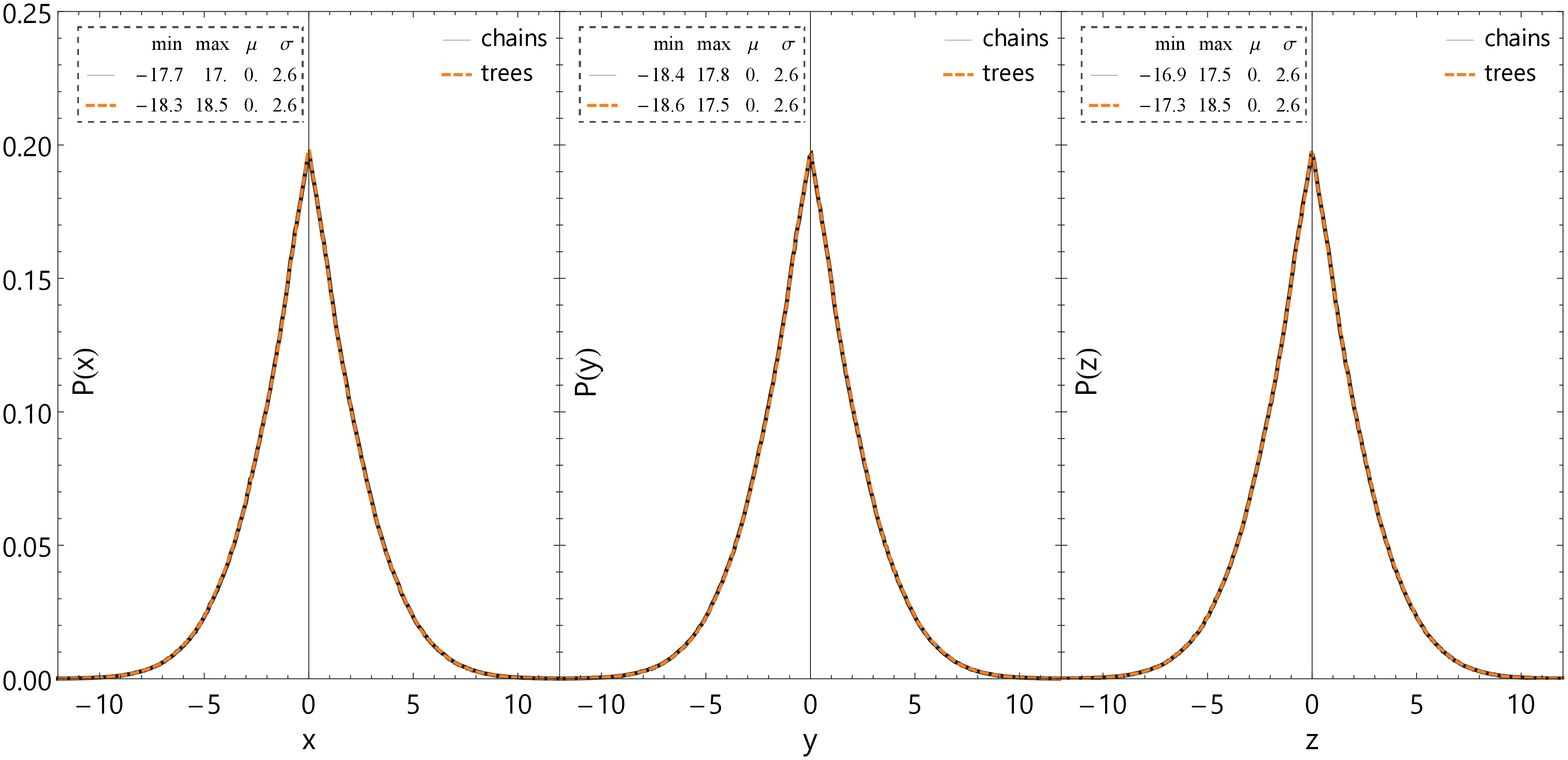}
 \caption{\emph{ The distributions of polarizations  in  the ensemble of  spin-chains by axes, which
 are calculated using $\mathbb{NP}$ (beige curves) and $\mathbb{P}$ (black curves) algorithms.} }
\end{figure}
Now we will prove the \textbf{hypothesis}  on the example of numerical experiments.
The characteristic distributions and parameters of the $1D$ spin glass, which
is in the state of the statistical equilibrium will be calculated using two $\mathbb{NP}$
and $\mathbb{P}$ algorithms. It is obvious that the comparison of the simulation results of the
relevant distributions will allow us to prove or disprove the \textbf{hypothesis}.

For simulation of the problem, first of all we have to set the initial conditions in the form
of a large number of independent configurations (roots);
$\bigl\{\Omega_1^1=(\textbf{\emph{s}}_1^1,\textbf{\emph{s}}_1^1;J_{1,2}^1)_1,...\Omega_N^1=
(\textbf{\emph{s}}_1^1,\textbf{\emph{s}}_2^1;J_{1,2}^1)_N\bigr\}=\hat{\Omega}$
(see the two scheme on Fig. 3).

Stages of simulation  using of the algorithm $\mathbb{NP}$  are as follows (the left scheme on Fig. 3).
Using the initial data, $\hat{\Omega} $ we perform parallel calculations of all graphs $G_i(n)$ of
the ensemble $G_i(n)\subset \{G(n)\}_N$. Note that each of these graphs in
terms of  classical mechanics, represents itself the set of classical trajectories that
go out from  one initial value (\emph{root}). The database, which is obtained in result of
simulation using $\mathbb{NP}$ algorithm allows to construct the distributions of the basic
parameters of the statistically equilibrium ensemble.

The simulation using  $\mathbb{P}$ algorithm (the right scheme on Fig. 3), is performed in a
similar way,   but with the difference that in this case instead of the set of graphs $\{G(n)\}_N$
we \emph{grow} the set of strings $\{g(n)\}_N$. In this case in the each graph we choose only
one string as the \emph{representative}. Note, that  the string (branch)
$g_j(n)\subset \{G(n)\}_N$ we grow by  way of randomly selecting only one solution in each node.
As a result of parallel simulation of the set of strings, we get the database which allows to
construct all distributions of the statistically equilibrium ensemble, $\{G(n)\}_N$ with the
asymptotic accuracy $O(N^{-1/3})$.

We compared  results of numerical simulations on the example of the statistical ensemble,
$\{G(20)\}_{5\cdot 10^4}$ consisting of $5\cdot 10^4$ graphs by heights $20$ with the ensemble
$\{g(20)\}_{5\cdot 10^4}$, which consists from the $5\cdot 10^4$ strings of lengths $20$.
As can be seen from Fig. 6 and Fig. 7, in the limit of statistical equilibrium,  the distributions
of various parameters of the statistical ensemble that have  calculated using of two $\mathbb{NP}$ and
$\mathbb{P}$ algorithms coincide ideal.

Thus we have shown on the example of 1$D$ Heisenberg spin glass, that the $\mathbb{NP}$ hard problem with
given accuracy may be reduced to the $\mathbb{P}$ problem and respectively the \textbf{hypothesis} is proved.


\section{Partition function}
Now it is important  return to the definition of the basic object of statistical physics,
i.e., to the partition function.

As well known,   the  multiparticle classical system in the state of statistical equilibrium
in the configuration space is described by the partition function of type:
\begin{equation}
\hspace*{-2.2cm}
 Z(\beta)=\int...\int \exp\bigl\{-\beta \,\mathcal{H}(\{\textbf{r}\})\bigr\}
 d\textbf{r}_1...,d\textbf{r}_N,\quad \beta=1/k_BT,\quad \{\textbf{r}\}=
 (\textbf{r}_1,...,\textbf{r}_N),
 \label{28}
\end{equation}
where $\mathcal{H}(\{\textbf{r}\})$ is the Hamiltonian of the system in the configuration
space, $k_B$ and $T$ are the Boltzmann constant and temperature of the system respectively.

For the considered model the partition function is calculated exactly \cite{Tomp}:
\begin{equation}
 Z(\beta,\{J\})=\prod_{i=1}^{n}\frac{sinh{(a_i)}}{a_i},\qquad a_i=\beta J_{i,\,i+1},
 \label{29}
\end{equation}
where the coupling constants; $J_{i,\,i+1}\in\{J\}=(J_{1,\,2},J_{2,\,3},...J_{n-1,\,n})$
are a random variables.

The average value of the partition function for the ensemble may be found by averaging over
the distribution of the coupling constant. Note that often assumed that this distribution is Gaussian:
\begin{equation}
W(J)=\frac{1}{\sigma_J\sqrt{2\pi}}
\exp\Bigl\{-\frac{(J-J_0)^2}{2\sigma_{J}^2}\Bigr\},
\label{30}
\end{equation}
where $\sigma_{J}$ is the variance and $J_0$ is the average value of coupling constant.

After averaging of  the expression (\ref{29}) by the distribution (\ref{30}) it is easy to find:
\begin{equation}
\hspace*{-2.6cm}
\bar{Z}(\beta)=\int_{-\infty}^{+\infty} Z(\beta,\{J\})W(J)dJ=\frac{K(\beta)}{\sqrt{2\pi}}
\int_{-\infty}^{+\infty}\Bigl(\frac{\sinh(\sigma_J\beta x)}{\sigma_J\beta x}\Bigr)^n \exp\Bigl\{
-\frac{1}{2}(x-x_0)^2\Bigr\}dx,
\label{31}
\end{equation}
where $x={J}/{\sigma_J}$ and $x_0={J_0}/{\sigma_J}$, in addition $K(\beta)$ denotes the normalization
factor of the partition function:
$$
K^{-1}(\beta)=\frac{1}{2\bar{J}}\int_{-\bar{J}}^{+\bar{J}}\Bigl(\frac{\sinh(J\beta )}{J\beta}\Bigr)^n dJ=
\frac{1}{\bar{y}}\int_{0}^{\bar{y}}\Bigl(\frac{\sinh(y)}{y}\Bigr)^n dy,\quad \bar{y}=\bar{J}\beta,
\quad J\in [\bar{J},-\bar{J}],
$$
 Recall that the coefficient $K(\beta)$ is constructed in such way that the Helmholtz free energy
in the limit $ \beta \to \infty $ converges to zero.

The Helmholtz free energy per one spin in chain is calculated as follow:
\begin{equation}
 F(\beta)=-\frac{1}{n\beta}\ln \bar{Z}(\beta).
 \label{32}
\end{equation}
\begin{figure}
\center
\includegraphics[height=65mm,width=70mm]{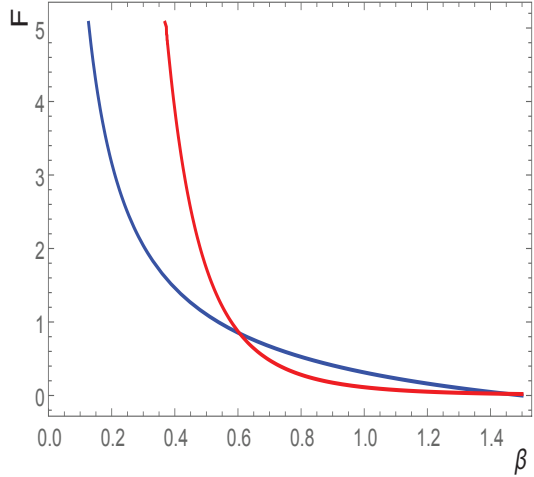}
\caption{\emph{The free energy of the ensemble which is calculated by two methods. The red curve
is obtained at using of the expression (\ref{33}), while the blue curve is obtained in the
result of calculation by the expression (\ref{32}). Note that parameters of $\varepsilon_0$
and $\sigma_\varepsilon$ are found by the way of simulation of problem from first principles,
whereas parameters $J_0$ and $\sigma_J$ chosen  on the basis of the best approximation to
the red curve.}}
\end{figure}
Since the integration in the representation (\ref{28}) is carried out by the full configuration
space, then obviously in such way taken into account  also  contributions of spin configurations,
which physically are unrealizable. Let us note that usually, the measure of set of such  spin
configurations is assumed to be equal to zero without  any serious proof, that not only groundlessly
but in a number of cases may be incorrect. Taking into account the fact that the set of strings
describing the statistical ensemble in configuration space formally can be represented as a
trajectory of dynamical system, in the limit of ergodicity of system (see [27, 28]), for
the partition function the following representation may be written:
\begin{equation}
 Z_\star(\beta)=\int_{-\infty}^{-n/\beta} \bar{P}(\varepsilon)
d\varepsilon,\qquad \bar{P}(\varepsilon)=c^{-1}P(\varepsilon),\quad
 c=\int_{-\infty}^{\,0}P(\varepsilon)d\varepsilon,
 \label{33}
\end{equation}
where $\varepsilon< 0$ denotes the energy of $1D$ spin-chain, while $\bar{P}(\varepsilon)$ is
the normalized  distribution of energy of the ensemble. Recall that $-n/\beta $ denotes
the limit of energy, above which there can be no stable spin-chain. As for the lower limit
then it should be $-\infty$, since on idea for all negative values of energies  exist a stable
spin-chain configurations,  nevertheless as seen from Fig. 6, starting from the value $\varepsilon_0=\mu$
(the average value of the spin-chain energy)  with decreasing of the energy the probability of
formation of spin-chains decreases.

 If the energy distribution (see Fig. 6)
to approximate by the Gaussian function (see (\ref{30})) then using the representation (\ref{33}),
for the free energy attributable to a single spin can be found the following expression:
\begin{equation}
F_\star(\beta)=-\frac{1}{n\beta}\ln\Bigl\{\frac{1}{2}\Bigl[1-erf\Bigl(\frac{\varepsilon_0+
n/\beta}{\sqrt{2}\sigma_\varepsilon}\Bigr)\Bigr]\Bigr\},
  \label{34}
\end{equation}
where $\varepsilon_0=\mu <0$ (see Fig. 6) denotes the average energy of spin-chain in the
ensemble and $\sigma_\varepsilon$, respectively, denotes the variance of spin-chains' energy
distribution. Comparing  Helmholtz's free energies,  $F(\beta)$ and $F_\star(\beta)$  for the
ensemble $\{g(20)\}_{5\cdot10^5}$ shows, that already at finite temperatures these curves diverge
significantly (see Fig. 8 ).  Furthermore, near the temperature $\beta\simeq 0.3 $, the  ensemble
of spin-chains exhibits a critical behavior, since the free energy tends to infinity that is
characteristic at phase transitions of first order. The latter obviously connected with  taking
into account of contribution non-physical configurations in the representation (\ref{28}), and
in formulas (\ref{29}) and (\ref{30}) respectively.

\section{Conclusion}

We have studied $1D$ spin glass in the framework of Heisenberg's nearest-neighboring Hamiltonian
(\ref{01}).  Using (\ref{01}) we obtained the system of recurrent algebraic equations (\ref{03}),
which together with conditions of energy minimum in nodes (\ref{05}) allow to implement node-by-node
calculations and to construct stable spin-chains. It is proved, that in the considered model, the
system of spins form only such spin-chains where all spins lie in one plane, while these planes
relative to each other may have any angle. Another important feature of the system of equations
(\ref{03}) consists in that there are probability of branching of solution in each node of 1$D$
lattice. This leads to the fact that in result of consecutive calculations,
from the one initial condition (\emph{root}) on the $n$-\emph{th} step, we get a set of
solutions (stable spin-chains or Kolmogorov's strings  $g_i(n)$) that form the \emph{Fibonacci
subtree} (random graphs $ G_j(n)\supseteq g_i(n)$). In other words, when we say on the
statistical ensemble we mean the set of random graphs $\{G(n)\}_N$, where $N$ denotes number
of graphs in the ensemble and correspondingly the problem consists in that to calculate all
parameters and corresponding distributions characterizing the ensemble.

It is shown that the computational complexity of  arbitrary graph $G_j(n)$ is the $\mathbb{NP}$
hard problem of the order $2^nK_s(n)$, while  complexity of the ensemble, with  increasing number
of elements  is increases linearly, $\{G(n)\}_N$  is the $\sim2^nNK_s(n)$. The properties of random graphs
depending on their height  are studied  in detail (see Fig.s 4-5) by using $\mathbb{NP}$ algorithm
(see the left scheme on Fig. 3) and conditions at which the ensemble $\{G(n)\}_N$ is in the state of
the statistical equilibrium are formulated. We analyzed and proposed the hypothesis that the $1D$
spin glass in the limit of statistical equilibrium may be simulated by using $\mathbb{P}$ algorithm
(see the right scheme on Fig. 3). Let us note, that all theoretical results and predictions have been
confirmed with high accuracy in numerical experiments that have been performed using $\mathbb{NP}$
and $\mathbb{P}$ algorithms (see Fig.s 5-7). It is noteworthy that the simulation by  the algorithm
$\mathbb{P}$ not only ensures high precision but also allows  to find  distributions of all parameters
of the ensemble, including the distribution of a constant spin-spin coupling  (see Fig. 5).

In the work has been suggested a new representation for the partition function in the
form of one dimensional integral from the spin-chain's energy distribution (see the expression (\ref{33})).
We  compared the Helmholtz free energies, which was calculated by using the usual (\ref{32})  and
new (\ref{34}) representations. As it is shown (see Fig. 8), already at  finite
temperatures the corresponding curves significantly different, moreover near $\beta\sim 0.3$ the
ensemble of spin-chains demonstrates critical property, that usually occurs  at first order phase
transitions. This is obviously due the fact that in the formula (\ref{31}), only such spin configurations
are counted which satisfy to the basic principles of classical mechanics (see expressions (\ref{03})
and (\ref{05})).

Thus, the main advantages of developed approach are that we have received clear answers, to all
raised questions on the example of study  $1D$ spin glass from first principles of the classical
mechanics without using any additional assumptions.  We showed that in the limit of statistical
equilibrium (at ergodicity of the statistical system), the initial $\mathbb{NP}$ hard problem
is reduced to the $\mathbb{P}$ problem, that allows radically simplify  the simulation of spin glasses.

The ideas lying in the base of developed approach enough are universal and allow the generalization
of model for a multidimensional case and at presence of external fields \cite{ash1}.

Finally, a new formulation of the problem of spin glasses and disordered systems in general can be
very useful for study of a global problem, i.e the problem of reduction  $\mathbb{NP}$ to the $\mathbb{P}$.

 \section*{References}

\bibliographystyle{elsarticle-num}

\end{document}